\documentclass[preprint,aps]{revtex4}     % for final submission
\usepackage{latexsym}

\draft
\begin{document}

% uncomment the following two lines for two-column version
%\twocolumnfalse\endcsname

\title{Effects of Vacancies on Properties of Relaxor Ferroelectrics: \\ a  First-Principles Study}
 
\author{ L. Bellaiche$^{1}$, Jorge \'I\~niguez$^{2}$, Eric Cockayne$^{3}$  and  B.P. Burton$^{3}$}

\address{$^{1}$ Physics Department,
                University of Arkansas, Fayetteville, Arkansas 72701, USA \\
                $^{2}$ Institut de Ciencia de Materials de Barcelona (ICMAB--CSIC), 
                Campus de la UAB, 08193 Bellaterra (Barcelona), Spain\\ 
                $^{3}$ Ceramics Division, Materials Science and Engineering Laboratory, 
                     National Institute of Standards and Technology, Gaithersburg, MD 20899-8520, USA 
                       }

\date{\today}

\begin{abstract}

A first-principles-based model is developed to investigate the
influence of lead vacancies on the 
properties of relaxor ferroelectric Pb(Sc$_{1/2}$Nb$_{1/2}$)O$_{3}$ (PSN).  Lead
vacancies generate large, inhomogeneous, electric fields that reduce
barriers between energy minima for different polarization directions.
This naturally explains why relaxors with significant lead vacancy 
concentrations have broadened dielectric peaks at 
lower temperatures, and why lead vacancies smear properties in the
neighborhood of the ferroelectric transition in PSN.
We also reconsider the conventional wisdom that lead vacancies reduce
the magnitude of dielectric response.

\end{abstract}

\maketitle

%

% uncomment the following line for two-column version
%\vskip2pc]

%\vspace{2mm}

\narrowtext

\marginparwidth 2.7in
\marginparsep 0.5in

\section{Introduction} 

Relaxor ferroelectrics are intriguing materials that 
characteristically exhibit a broad frequency-dependent 
dielectric response ~\cite{Cross}. 
Although relaxors were discovered more than fifty years ago~\cite{Smolenskii}
the precise origin of their peculiar dielectricity remains controversial.
Various models have been proposed (e.g., Refs.  
\cite{Westphal,Tagantsev,Viehland,Colla,Pirc,Ben1,JorgeLaurentPSN} and references therein) 
all of which include quenched, randomly distributed, 
internal electric fields  and/or interacting polar microregions.

Experimentally, the introduction of 
vacancies into relaxors induces greater broadening of the dielectric response; 
i.e. it {\it enhances} relaxor character \cite{Chu1,Malibert,Chu2}. 
Other Pb-vacancy induced effects include: shifting of the dielectric peak
to lower temperature; a decrease in the magnitude of the dielectric response;
and a more ``diffuse'' phase transition  \cite{Chu1,Malibert,Chu2}.

The mechanisms responsible for vacancy-induced characteristics are mostly unknown, 
but understanding them would illuminate the fundamentals of relaxor
behavior, especially because defects are always present in
real samples.  One reason for this paucity of knowledge is that first-principles 
studies of (point) defects in ferroelectrics are scarce \cite{Eric,Poykko,Ben,Horace}.

In this manuscript, we investigate the effects of Pb-vacancies on the properties of  
a relaxor ferroelectric from first principles. 
Our model system is Pb(Sc$_{1/2}$Nb$_{1/2}$)O$_{3}$ (PSN) rather than the 
prototype relaxor Pb(Mg$_{1/3}$Nb$_{2/3}$)O$_3$ (PMN), because
the former is easier to model -- i.e. finite-temperature first-principles-based 
models already exist for PSN \cite{Ben1,APLPSN} but not for PMN.
Note that, unlike PMN which retains relaxor character down to 0K 
(i.e. exhibits no ferroelectric phase transition), 
chemically disordered PSN  exhibits  relaxor behavior 
between $\sim$ 400K and $\sim$ 368K, then transforms to 
a rhombohedral ferroelectric phase~\cite{Chu1,Malibert}.
We focus on static properties, because studying the dynamics of relaxors 
from first-principles is not currently feasible. Our study 
clarifies the origin(s) of the vacancy-induced effects mentioned above, 
and suggests that the common wisdom that Pb-vacancies reduce 
the magnitude of dielectric response should be reconsidered. 

This manuscript is organized as follows: Section II describes
the method.  Section III presents and discusses results.
Section IV is a brief summary.

\section{Method} 

Our method is a generalization of the first-principles-derived
effective Hamiltonian of Refs.~\cite{APLPSN,PRLPZT,FEPZT} to investigate
perovskites with chemical disorder on their B-sites and Pb-vacancies
on their A-sites, by writing the total energy as a sum of two terms:

\begin{eqnarray}
   E_{\rm PSN-DV} (\{ { \bf u_{\it i}} \}, \eta_{\it H}, \{  \eta_{\it I} \},  \{ \sigma_{\it j} \}, \{ s_{\it k} \})
       \nonumber \\
   =~ E_{\rm PSN-D} (\{ { \bf u_{\it i}} \}, \eta_{\it H},\{ \eta_{\it I}\},\{ \sigma_{\it j} \}) 
  +~E_{\rm vac} (\{ { \bf u_{\it i}} \}, \{ s_{\it k}  \} ) \;\;,
\end{eqnarray}

where $E_{\rm PSN-DV}$ is the total energy of the 
Pb$_{1-x} \Box _{x}$(Sc$_{1/2}$Nb$_{1/2}$)O$_3$ alloy (PSN-DV),
$\Box$~representing a vacant A-site; 
$E_{\rm PSN-D}$~ is the total energy of chemically disordered PSN solid solutions (PSN-D);
$E_{\rm vac}$  gathers the explicit energetics associated with Pb-vacancies; 
${\bf u_{\it i}}$ is the (B-centered) local polar distortion in unit cell $i$ 
(proportional to the local dipole moment);
$\eta_{\it H}$ and $\{  \eta_{\it I} \}$ are the {\it homogeneous} 
and {\it  inhomogeneous} strain tensors \cite{ZhongDavid}, 
respectively;   $\sigma_j = -1$ or $+$1 if there 
is a Sc or Nb cation, respectively, 
at the B-lattice site $j$ of the Pb$_{1-x} \Box _{x}$(Sc$_{1/2}$Nb$_{1/2}$)O$_3$ alloy.  
Finally, $ \{ s_{\it k}  \} $  characterizes the amount and distribution of 
A-site vacancies, i.e. $s_{\it k}$=$0$ or $1$ if the $k$'th A-site is 
occupied by a lead atom or vacant, respectively. 

For $E_{\rm PSN-D}$, we use the analytical expression proposed in 
Ref. \cite{APLPSN}, which includes a term of the form 
$ - \sum_i Z^* {\bf u}_i \cdot {\bf \epsilon_i}[\{ \sigma_{\it j} \}] \ $, 
where $Z^*$ is the Born effective charge associated with the local 
distortion, and ${\bf \epsilon}_i$ is the internal field at cell $i$ 
that is caused by the surrounding B- site cation distribution~\cite{AaronJorge}.

For  $E_{\rm vac}$, we use the following pertubative expression 
(first-order in ${\bf u_{\it i}}$): 

\begin{equation}
  E_{\rm vac}  (\{ { \bf u_{\it i}} \}, \{ s_{\it k}  \} )=  \sum_{k} \sum_{i} S_{\it k,i}~ s_{\it k} ~ { \bf g_{\it ki}} \cdot { \bf u_{\it i}} 
\end{equation}

where the sum over $k$ runs over A-sites, and the sum over $i$ runs over B-sites. 
The $S_{{\it k,i}}$ parameters quantify how vacancies perturb the local distortions. 
$\bf{g_{\it ki}}$ is the unit vector joining the A-site $k$ to the B-site-center 
of $\bf{u_{\it i}}$.  We truncated $S_{\it k,i}$ 
at the first-neighbor shell.  All the parameters of Eqs.~(1) and (2) are 
derived from first-principles calculations \cite{USPP,LDA,LaurentDavid3} performed 
on relatively small supercells (i.e., up to 40 atoms), 
including one (to obtain $S_{{\it k,i}}$) that exhibits a lead vacancy and a background
charge of $-$2. 

We solve this effective Hamiltonian by the Monte Carlo (MC) technique \cite{Metropolis}
to simulate properties of two different systems: 
1) disordered Pb(Sc$_{1/2}$Nb$_{1/2}$)O$_{3}$~
{\it without any Pb-vacancies}, PSN-D; 
2) disordered Pb$_{0.95}\Box_{0.05}$(Sc$_{1/2}$Nb$_{1/2}$)O$_{3}$,
PSN-DV, which has a realistic 5\% vacancy concentration on its A-sites ~\cite{Chu1,Malibert}). 
A $12\times 12\times 12$ supercell (8640 atoms, 48 \AA ~ lateral size) 
is used for both systems. 
The variables $\sigma_j$ and $s_{\it k}$ are randomly chosen, and remain constant
during the simulation.  Temperature, $T$, is decreased in small steps from high $T$.
We found it unusually difficult to get good statistics and, therefore, performed 10 MC
runs of 4 {\sl million} sweeps each, at every temperature.  
We collect the supercell-average ${\bf u}$ of the local distortion at each sweep and classify 
it as {\sl visiting} one of the eight rhombohedral, twelve orthorhombic, or six tetragonal 
energy minima by computing its projection onto each of these twenty-six high-symmetry directions. 
The {\sl visiting rate} of a particular minimum is defined as the ratio between
the number of sweeps in which that minimum is visited and the total number of
sweeps. The $\chi_{\alpha \beta}$ dielectric susceptibility tensor is computed as in Refs.~\cite{Alberto1,Karin}, that is by using the polarization fluctuation formula: 
\begin{eqnarray}
\chi_{\alpha \beta}= \frac{(N~Z^{*})^2}{V\epsilon_o k_BT} \left [ < u_{\alpha} u_{\beta} > - < u_{\alpha}> < u_{\beta} >  \right ]
\end{eqnarray}

where $< u_{\alpha} u_{\beta}>$ denotes the statistical average of the product between the  $\alpha$ and $\beta$ components of  the supercell average of the  local mode vectors, and where $< u_{\alpha}>$  (respectively, $< u_{\beta}>$) is the statistical average of the $\alpha$- (respectively, $\beta$-) component of the supercell average of the  local mode vectors.
$N$ is the number of sites in the supercell while $V$ is its volume.  $k_B$ is the Boltzmann's constant and $\epsilon_o$ is the permittivity of the vacuum.
Note that a detailed derivation of Eq.(3) is provided in Ref. \cite{Karin}.

\section{Results} 

The simulated temperature at which the dielectric response is maximized, $T_m$, 
is about 850K for PSN-D and 794K for PSN-DV \cite{Footnoterot}.  
Thus, our model qualitatively reproduces the  
previous finding that Pb-vacancies reduce $T_m$ \cite{Chu1,Malibert,Ben}. 
Our quantitative prediction of a 6.5 \% reduction in T$_m$, from the introduction 
of 5\% Pb-vacancies agrees rather well with the corresponding experimental reduction of 
8.5\% \cite{Malibert}, and is also consistent with 
the  $\simeq$ 2.6\% reduction of  T$_m$ associated with 1.7\% Pb-vacancies 
\cite{Chu1}; establishing the accuracy of 
simulations based on Eqs. (1) and (2). 
Note that Eq.(2) can be rewritten as  
$E_{\rm vac}  (\{ { \bf u_{\it i}} \}, \{ s_{\it k}  \} )=  
- \sum_i Z^* {\bf u}_i \cdot {\bf \epsilon_{vac,i}}[\{ s_{\it k} \}] \ $, with  ${\bf \epsilon_{vac,i}} 
= - \frac{1}{Z^* }  \sum_{k}  S_{\it k,i}~ s_{\it k} ~ { \bf g_{\it ki}}$;
that is, ${\bf \epsilon_{vac,i}}$ has the dimensions and physical meaning of 
an internal electric field arising from the Pb-vacancies. 
Our first-principles-derived values for the fields around a Pb-vacancy
are of the order of 1.4~$\times 10^9$~V/m at each neighboring B-site $i$. 
These fields are oriented towards the vacancy because  
$S_{{\it k,i}}$ is positive and ${\bf \epsilon_{vac,i}}$  depends 
on the directional $ { \bf g_{\it ki} }$ vectors;  
${\bf \epsilon_{vac,i}}$ is therefore rather large and 
{\it inhomogeneous}, and thus significantely {\it opposes} 
homogeneous ferroelectric  $\{ { \bf u_{\it i}} \}$ ordering, 
which explains the vacancy-induced reduction of T$_m$.

Figure~1a displays the $(u_x, u_y, u_z)$
{\it supercell} averages of local distortions as functions of the
reduced temperature $T/T_m$; [ $(u_x, u_y, u_z)$ is directly proportional
to the spontaneous polarization, where the $x$, $y$, and $z$ axes
are chosen along the pseudo-cubic [100], [010], and [001] directions,
respectively]. 
Both PSN-D and PSN-DV are predicted to have a low-$T$
rhombohedral ferroelectric phase in which $u_x\approx u_y\approx u_z$,
in agreement with experiment \cite{Chu1,Malibert}. Therefore, only the 
average component, 
$\langle \bar{u}\rangle=1/3\langle u_x + u_y + u_z \rangle$, is plotted in Fig.~1a. 
Also shown in Fig.~1a is $\langle |u|\rangle_R/\sqrt{3}$, where $\langle
|u|\rangle_R$ is the mean {\sl magnitude} of the supercell average of
local distortions calculated from sweeps in which the system visits
rhombohedral minima. Figure~1b shows $\chi$ -- which corresponds to  one
third of the trace of the $\chi_{\alpha \beta}$ dielectric susceptibility tensor -- versus $T/T_m$. 
In other words, and according to Eq. (3) and the definition of $\chi$, Fig.~1b displays:

 \begin{eqnarray}
\chi= \frac{(N~Z^{*})^2}{V\epsilon_o k_BT} \frac{1}{3} 
\left [ < u_x ^2 +  u_y ^2 + u_z ^2 > - ( < u_x>^2 + < u_y>^2 + < u_z>^2 ) \right ]
\end{eqnarray}

Predictions for $\langle
\bar{u}\rangle$ and $\chi$ are quite scattered around
$T_m$, which reflects our difficulties in obtaining good
statistics for either PSN-D or PSN-DV, as expected for systems 
with complex energy landscapes \cite{Viehland}. 
We smoothed the results for $\langle \bar{u}\rangle$ by fitting them to a 7th-order
polynomial in the region around $T_m$ (lines in Fig.~1a). We denote
the result by $\langle \bar{u}\rangle_{\rm fit}$ and recompute the
susceptibility as
\begin{eqnarray}
\chi_{\rm fit} = \frac{N~(Z^{*})^2}{a_0 ^3 \epsilon_o k_BT}  (
\frac{1}{3}\langle |u|\rangle_R^2 - \langle \bar{u}\rangle_{\rm fit}^2)
\, ,
\end{eqnarray}
where  $a_0$ is the 0K cubic
lattice constant. We obtain Eq.~(5) from Eq.~(4), by approximating $V$ as $N a_0 ^3$, and the 
averages as follows: $ 1/3 < u_x ^2 +  u_y ^2 + u_z ^2 >  \sim 1/3\langle |u|
\rangle_R^2$ and $ 1/3 (< u_x>^2 + < u_y>^2 + < u_z>^2)   \sim \langle
\bar{u}\rangle_{\rm fit}^2$ \cite{FootnoteCorrel}. The ``smoothed'' $\chi_{\rm fit}$'s are
depicted with lines in Fig.~1b, and will be used 
in combination with $\langle \bar{u}\rangle_{\rm fit}$,
to illuminate the effects of Pb-vacancies in relaxors.

Interestingly, Fig.~1b indicates that, contrary to common wisdom and 
previously reported observations \cite{Chu1,Malibert}, our predicted 
maximum value of the dielectric response does {\it not} decrease 
when Pb-vacancies are introduced: our simulations indicate that
$\chi_{\rm fit}$ is slightly larger in PSN-DV than in PSN-D at $T=T_m$, because
T$_m$ is smaller and $\langle \bar{u}\rangle_{\rm fit}$ is further away from
$\langle |u|\rangle_R/\sqrt{3}$ in PSN-DV, see Fig. 1a and Eq.~(5).
[Our predicted maximum values for $\chi_{\rm fit}$ are about 34,000 and 38,500 
for PSN-D and PSN-DV, respectively, which  is close to the measured 
value $\simeq$ 40,000 of Ref. \cite{Chu1} for PSN-D].
This result, which appears to be at odds with measurements, is 
explained by the experimental observation of Ref. \cite{Stage} that 
Pb-vacancies in Pb(Sc$_{1/2}$Nb$_{1/2}$)O$_{3}$ favor the simultaneous 
formation of pyrochlore phases (Pb$_{2.31}$Nb$_2$O$_{7.31}$ and Pb$_3$Nb$_4$O$_{13}$). 
Thus, dielectric measurements on a PSN-DV sample at $T=T_m$ should yield a {\it smaller} 
response than those on PSN-D; because, while the majority of the PSN-DV sample
is perovskite, with a dielectric response that is at least as large as that for PSN-D, 
a smaller fraction is pyrochlore, which has a much {\it smaller} 
dielectric response \cite{Stage}.

Figure 1b also shows that our simulations successfully 
reproduce two experimental observations:
(i) the {\it broadened} dielectric peaks in PSN-D and PSN-DV, 
which are characteristic of relaxors~\cite{Cross}, and according
to Eq. ~(5) and Fig~1a,  arise from the diffuse phase transition associated 
with $\langle \bar{u}\rangle_{\rm fit}$; (ii) Pb-vacancy induced 
dielectric {\it broadening} \cite{Chu1,Malibert,Chu2}.
Half-maximum width of the $\chi_{\rm fit}$ peak is about 0.19 in PSN-D
{\sl versus} 0.22 in PSN-DV (in reduced units)  [c.f. the corresponding
predicted value of 0.13 reported in Ref. \cite{JorgeLaurentPSN} for the ``normal'' 
ferroelectric system Pb(Zr$_{0.6}$Ti$_{0.4}$)O$_3$]. 
Note that vacancy-induced dielectric broadening only occurs {\it below} T$_m$.
This is particularly clear in the Fig. 1b inset, in which the 
$\chi_{\rm fit}$~ of PSN-DV is rescaled by a T-independent constant to
match the $\chi_{\rm fit}$~ dielectric response of PSN-D at $T=T_m$.

To clarify how PSN-DV differs from PSN-D only at temperatures 
below T$_m$, Table I lists the internal energies of rhombohedral and orthorhombic minima 
(calculated at 5\,K); plus the visiting rates of these minima in both PSN-D 
and PSN-DV at $T/T_m \simeq 0.93$.
The energy landscape of PSN-D is rather anisotropic; e.g. the energy difference between
the ground state and the highest-lying rhombohedral minimum is 0.68~meV. 
Reference \cite{JorgeLaurentPSN} demonstrated that such local symmetry breaking is caused
by the anisotropy of the large internal ${\bf \epsilon}_i$ electric fields 
(which have an average magnitude $\simeq$ 2 $\times 10^9$~V/m; determined numerically)
that arise from the charge-difference between Sc$^{3+}$~ and Nb$^{5+}$~ ions. 
This anisotropy results from our nanometric 
periodic PSN-D supercell 
being too small to contain a perfectly random B-site configuration \cite{JorgeLaurentPSN}. 
Results in Table I also indicate that, at $T/T_m \simeq 0.93$, PSN-D preferentially visits the 
rhombohedral [-111] and [111] minima, i.e. the minima that have {\sl lowest} energies, 
and that are connected by one of the lowest-in-energy 
orthorhombic minima (specifically, [011]). Such fluctuations lead to a 
$\langle \bar{u}\rangle_{\rm fit}$ that is smaller than
$\langle |u|\rangle_R/\sqrt{3}$, and thus [Eq.~(5)], to a large dielectric response 
relative to normal ferroelectrics (which are already stuck in a single minimum 
at $T/T_m \simeq 0.93$ \cite{JorgeLaurentPSN}).
As indicated by Table I, the ${\bf \epsilon_{vac,i}}$ electric field associated 
with Pb-vacancies (see rewriting of Eq~(2) above) 
brings the orthorhombic and rhombohedral minima closer to each other; similar to the finding
of Ref. \cite{JorgeLaurentPSN} that ${\bf \epsilon}_i$ in PSN-D reduces 
barriers between energy minima. 
Thus, PSN-DV fluctuates more and/or visits more minima than
PSN-D at temperatures below T$_m$. This implies a smaller 
$\langle \bar{u}\rangle_{\rm fit}$  (Fig.~1a) and therefore naturally 
explains two vacancy-induced 
effects that have been observed experimentally \cite{Chu1,Malibert,Chu2}:
a broader PSN-DV dielectric response  (Fig.~1b and Eq. (5)); 
and a more diffuse PSN-DV phase transition 
(PSN-DV ``only'' begins to stick in a single rhombohedral minimum
at $T/T_m \simeq 0.85$ as indicated by the equality
$\langle \bar{u}\rangle_{\rm fit} = \langle |u|\rangle_R/\sqrt{3}$ 
at and below this temperature, Fig~1a, 
whereas, freezing in of PSN-D occurs at $T/T_m \simeq 0.90$).

Results in Table I also indicate that the most frequently visited rhombohedral minima 
in PSN-DV are {\it not} the lowest-energy minima; rather they are those with small-in-energy 
rhombohedral-to-orthorhombic 
barriers (e.g., the [1-1-1] and [-1-1-1] minima that are connected 
by the [0-1-1] minimum, over an energy barrier of  about 2.5 meV). 
This  strongly suggests that relaxors with Pb-vacancies can be 
thought of as non-ergodic systems \cite{Igornonergodic}.

\section{Summary} 

In summary, first-principles-based calculations predict that  the introduction 
of Pb-vacancies in Pb(Sc$_{1/2}$Nb$_{1/2}$)O$_{3}$: (i) reduces the temperature 
at which the dielectric response peaks, $T_{m}$~, consistent with 
experiment \cite{Chu1,Malibert,Chu2}; 
(ii) does {\it not intrinsically } decrease  the magnitude of the dielectric peak, 
contrary to common wisdom \cite{Chu1,Malibert,Chu2}; 
(iii) broadens the dielectric response and enhances the diffuse character 
of the ferroelectric transition, consistent with measurements  \cite{Chu1,Malibert,Chu2} 
(but only at temperatures below T$_m$). Item (i) above
is the direct consequence of the relatively large 
internal inhomogeneous electric fields generated by the Pb-vacancies. 
We suggest that our disagreement 
with experiments in item (ii) is caused by parasitic pyrochlore phases 
in the experimental samples. Item (iii) originates from the decrease 
in energy barriers that are caused by Pb-vacancies, which promote 
enhanced fluctuations between different energy-minima below T$_m$.
Interestingly, our results regarding the effect of defects on energy landscape may also be 
relevant for modeling (and understanding) {\it dynamical} properties of relaxors.

{\bf Acknowledgment} 

We thank J.M. Kiat, B. Dkhil, and C. Malibert for useful
discussions. This work is supported by ONR grants N00014-01-1-0365,
N00014-04-1-0413 and N00014-01-1-0600, by NSF grant DMR-0404335, and
by DOE grant DE-FG02-05ER46188. JI thanks financial support from the
Spanish Ministry of Science and Education (``Ram\'on y Cajal'' program
and grant BFM2003-03372-C03-01), the Catalan Government grant
SGR-2005-683, and FAME-NoE.

%\newpage
 %
\begin{table}
\caption{Total energy per five-atom cell $E$, at 5K, and visiting rate $v_r$ at $T/T_m \simeq 0.93$ associated with  the different rhombohedral $\langle111\rangle$  and orthorhombic
 $\langle110\rangle$ minima in PSN-D and PSN-DV. The zero of energy corresponds to the ground-state.  The uncertainty is lower than 0.01 meV and 0.1\% for $E$ and $v_r$, respectively.}

 \begin{tabular}{ccccccccc}
  \hline 
Minimum & & E (meV) in PSN-D & & E(meV) in PSN-DV & & $v_r$ in PSN-D (\%) & &  $v_r$ in PSN-DV (\%) \\
\hline
[-1-11] & & 0.38 & & 0.41 & & 6.3 & &  0.0 \\
 \hline
 [-111] & & 0.00 & & 0.23 & & 72.9 & &  0.0 \\
\hline
 [111] & & 0.07 & & 0.00 & & 20.0 & &  0.0 \\
\hline
 [1-11] & & 0.68 & & 0.58 & & 0.0 & &  0.0 \\
\hline
[1-1-1] & & 0.00 & & 0.28 & & 0.0 & &  37.2 \\
\hline
[11-1] & & 0.67 & & 0.33 & & 0.0 & &  17.3 \\
\hline
[-11-1] & & 0.65 & & 0.38 & & 0.0 & &  5.8 \\
\hline
[-1-1-1] & & 0.20 & & 0.32 & & 0.0 & &  37.5 \\
\hline
\hline
[-101] & & 3.05 & & 2.90 & & 0.3 & &  0.0 \\
\hline
 [011] & & 3.05 & & 2.80 & & 0.5 & &  0.0 \\
\hline
 [101] & & 3.45 & & 2.85 & & 0.0 & &  0.0 \\
\hline
[1-10] & & 3.42 & & 3.42 & & 0.0 & &  0.0 \\
\hline
[10-1] & & 3.48 & & 2.93 & & 0.0 & &  0.7 \\
\hline
[01-1] & & 3.85 & & 2.99 & & 0.0 & &  0.2 \\
\hline
[-10-1] & & 3.60 & & 3.12 & & 0.0 & &  0.2 \\
\hline
[0-1-1] & & 3.05 & & 2.85 & & 0.0 & & 1.1 \\
\hline
[0-11] & & 3.71 & & 3.29 & & 0.0 & &  0.0 \\
\hline
[110] & & 3.69 & & 3.34 & & 0.0 & &  0.0 \\
\hline
[-110] & & 3.54 & & 3.31 & & 0.0 & &  0.0 \\
\hline
[-1-10] & & 3.58 & & 3.53 & & 0.0 & &  0.0 \\
\hline
\hline

\end{tabular}
\end{table}

\newpage

\begin{figure}
\caption{Properties of PSN-DV (open dots, dashed lines) and PSN-D (filled
{dots, solid lines) as functions of $T/T_m$. Dots in panel~(a) show the 
supercell-averaged mean component of the local distortions ($\langle \bar{u}\rangle$). 
Lines represent fits to a 7th order polynomial
($\langle \bar{u}\rangle_{\rm fit}$). In panel (a), we also show (squares) the
magnitudes of the local distortions, $\langle |u|\rangle_R$, divided by
$\sqrt{3}$. Panel (b) displays one
third of the trace of the dielectric susceptibility tensor directly
obtained from MC simulations. The lines represent the $\chi_{\rm fit}$ results
obtained from Eq.~(5)}. The inset of Panel (b) displays $\chi_{\rm fit}$ results 
that were rescaled to match $\chi_{\rm fit}$ of PSN-D at $T=T_m$. }
\end{figure}


\begin{references}
\bibitem{FootnoteCorrel} Note that we numerically checked that 
(i) $ 1/3 < u_x ^2 +  u_y ^2 + u_z ^2 >$ is very well approximated by  $1/3\langle |u|
\rangle_R^2$;  (ii)  $ 1/3 (< u_x>^2 + < u_y>^2 + < u_z>^2)$ is also well approximated by $  \langle
\bar{u}\rangle_{\rm fit}^2$; and (iii) the polarization fluctuation formula [see Eq.~(3)] provides the same dielectric constant than the one (directly) obtained by computing the electric-field-induced change in polarization.
\end{references}
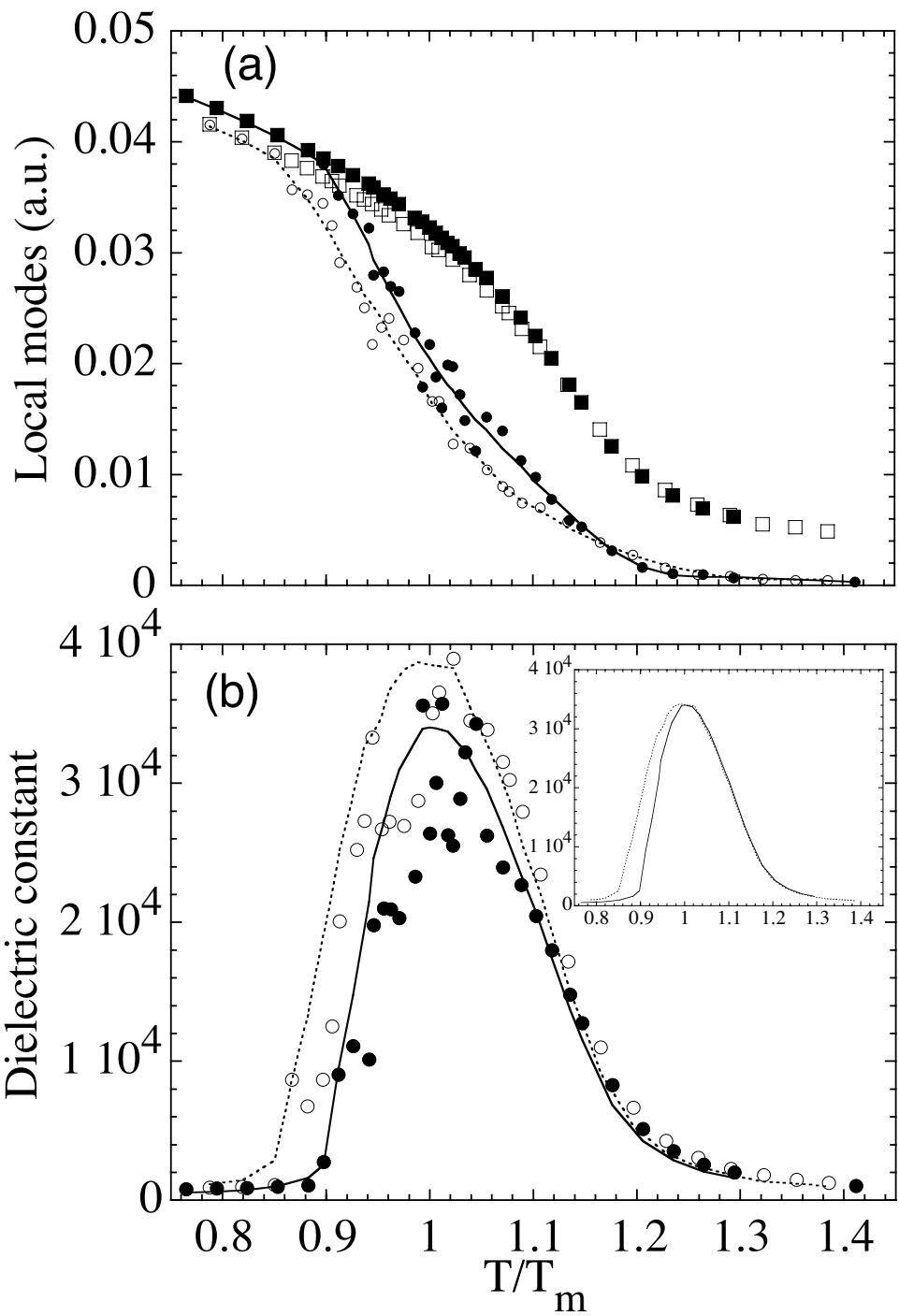
\end{document}